\documentclass[preprint,showpacs,preprintnumbers,amsmath,amssymb]{revtex4}
\usepackage{graphicx}
\bibpunct{[}{]}{;}{a}{,}{,}

\begin{document}

\title{Local enrichment and its nonlocal consequences for victim-exploiter
metapopulations}

\author{Gur Yaari$^{1,2}$,  Sorin Solomon$^{2,1}$, Marcelo
Schiffer$^{3}$ and Nadav M. Shnerb$^{4}$}

\affiliation{($1$) Multi-Agent Systems Division, Institute for Scientific Interchange Foudation, 10133 Turin, Italy\\ ($2$)Racah Institute of Physics, The Hebrew University, Jerusalem 91910 Israel \\ ($3$)
Department of Physics, Judea and Samaria College, Ariel 44837
Israel.\\ ($4$)  Department of
Physics, Bar-Ilan University, Ramat-Gan 52900 Israel  }

\pacs{87.17.Aa,05.45.Yv,87.17.Ee,82.40.Np}

\begin{abstract}
The stabilizing effects of local enrichment are revisited.
Diffusively coupled host-parasitoid and predator-prey
metapopulations are shown to admit a stable fixed point,  limit
cycle or  stable torus with a rich bifurcation structure. A linear
toy model that yields many of the basic qualitative features of this
system is presented. The further nonlinear complications are
analyzed in the framework of  the  marginally stable Lotka-Volterra
model, and the  continuous time analog of the unstable,
host-parasitoid Nicholson-Bailey model. The dependence of the
results on the migration rate and level of spatial variations is
examined, and the possibility of "nonlocal"  effect of enrichment,
where local enrichment induces stable oscillations at a distance, is
studied. A simple method for basic estimation of the relative
importance of this effect in experimental systems is presented and
exemplified.
\end{abstract}

 \maketitle

\section{Introduction}

Spatially extended ecological systems, dispersal and metapopulation
dynamics have attracted a lot of interest in recent years
\cite{Kariva_Tilman1997}. In particular, the stability of an
ecological system that contains many species interacting  via
competition, predation and symbiosis, and subject to environmental
and demographic noise,  poses an interesting mathematical and
ecological puzzle. The role of dispersal among sites, migration
amongst  habitat patches, "rescue" of a habitat by dispersal from
other locations and so on, have been recognized as  crucial
ingredients in the stabilization mechanism.

This work concentrated on one of  the basic  ecological processes -
the dynamics of  victim-exploiter (predator-prey, host-parasitoid)
populations.  In the "mean field" limit, i.e., where
 the chance for a predation event is fixed for any exploiter-victim pair, the simplest
mathematical models are those of  Lotka and Volterra
\cite{Lotka1920, Volterra1931} for predator-prey,  and
Nicholson-Bailey \cite{Nicholson_Bailey1935} for host-parasitoid
populations.  Both models do not allow for an attractive manifold
(such as  fixed point, limit cycle or strange attractor). In fact,
the coexistence point is either marginally stable (Lotka-Volterra
like) or unstable (with the Nicholson-Bailey host parasitoid system
as a prototype). In both cases one may expect to see extinction of
(at least) one of the species after a short time; if the system is
marginally stable the noise will drive it to extinction, while for
an unstable manifold the system gravitates perpetually towards the
edge of extinction.

How do victim-exploiter systems persist in nature for millions of
years? One may suggest that the simple models are wrong, and assert
that any realistic mathematical description of a victim-exploiter
system should be "dissipative", supporting an attractive manifold.
There are many modifications of the basic models that achieve these
goals (e.g., by taking into account the finite carrying capacity of
the environment or  adding delayed response), and  the observed
population oscillations may be  a result of noise perturbing an
attractive fixed point \cite{McKane_Newman2005}. However, since the
work of Gause \cite{Gause1934} through the classic experiments of
Pimentel et al. \cite{Pimentel_ea1963}, Luckinbill
\cite{Luckinbill1974} and Huffaker \cite{Huffaker1958}, it was known
that small sized  predator prey (or hosts and parasites) systems
reach extinction in experimental time scales. The reason for this
has become clear in the last decade, due to the experiments of
Holyoak and Lawler \cite{Holyoak_Lawler1996}, Kerr et al.
\cite{Kerr_ea2002,Kerr_ea2006} and Ellner et al.
\cite{Ellner_ea2001}. In all of  these experiments the same system
goes extinct rapidly in the well-mixed limit while persisting way
above the experimental time (up to h undreds of generations) when
the population is spatially segregated. It follows that in many
cases, and perhaps even generically, victim-exploiter systems are
\emph{unstable} in the well mixed limit, and acquire their stability
due to its spatial structure.

What exactly stabilizes the spatial structure?  a few answers have
already been suggested. Some are related to the effect of spatial
heterogeneity or spatio-temporal fluctuations, while others stress
more generic mechanisms that may yield stability even if the space
is perfectly homogenous. In this paper we focus on the effect of
spatial heterogeneity (e.g., different growth rates on different
spatial patches)  as a stabilizer.

The  observation that diffusive coupling between patches may
stabilize otherwise unstable dynamics, or may result in convergence
to a focus instead of a limit cycle, has been made in few
disciplines independently. In ecology, Murdoch and Oaten
\cite{Murdoch_Oaten1975} suggested that dispersal between
Lotka-Voltera patches with spatial variability may stabilize the
fixed point. Subsequent  works by Crowley \cite{Crowley1981}, Ives
\cite{Ives1992}, Murdoch et. al. \cite{Murdoch_ea1992} and Taylor
\cite{Taylor1998}, extended  this basic idea to include multi-patch
systems,  the effects of parasitoid aggregation, differences in
diffusion parameters, density different migration and other
complications that may occur in realistic systems. In chemistry,
this stabilization is known alternatively as "oscillator death" and
was observed by Bar-Eli \cite{Bar-Eli1985} in the context of coupled
chemical oscillators. That basic idea has since been applied to
other diffusively coupled chemical systems, such as neural
\cite{Kopel_Ermentrout1990} and calcium oscillations
\cite{Tsaneva-Atanasova_ea2005}. Mathematically speaking, the
stabilizing effect of diffusive coupling between two sites on a
single species, extinction prone chaotic system has been considered
by Gyllenberg et. al. \cite{Gyllenberg_ea1996}. These authors
pointed out the "salvage effect", where the existence of a sink may
stabilize the population on the source habitat; this effect also
appears in the systems considered below.

Although the basic phenomenon is known, it turns out that the system
of diffusively coupled unstable oscillations is quite rich and may
reveal  many  interesting features beyond the stabilization of a
fixed point.  In this paper we   analyze a very simple case of a
single enriched site (where the prey, or host population flourish)
surrounded by a less productive environment. We will examine in
detail  the conditions for the stability of the fixed point and the
asymptotic behavior of the Lyapunov exponent, consider the
possibility for the appearance of a limit cycle, and discuss the
associated bifurcations. The most interesting phenomenon observed
relates to the spatial population profile: here the effect of
localized enrichment  may yield either local or nonlocal changes,
including the emergence of oscillations far from the location of
enrichment, incommensurate oscillations, etc.

This paper is organized as follows:  In the next section a linear
toy model is presented and a few basic insights are derived. The
effect of nonlinearities is emphasized for a predator-prey,
marginally stable model in the third section. Although marginal
stability is not a robust feature of a dynamical system and should
not be considered seriously as the underlying dynamic of ecological
processes, its  consideration enables the attainment of basic
knowledge regarding  the effect of nonlinearity on  stability, as
will explained later on. The fourth section analyzes a
Nicholson-Bailey like (unstable) system  and discusses  interesting
nonlocal effects of enrichment. Finally, our results are discussed
in view of the "paradox of enrichment" \cite{Rosenzweig1971}, and
possible experimental tests are suggested.

\section{A toy model: coupled linear  oscillators}

Let us present a linear model in order to exemplify  some features
of the fixed point stability (local analysis) for non-identical
victim-exploiter patches. We consider two diffusively coupled
harmonic oscillators (with the possibility of  repulsive term with a
positive constant  $\alpha$):

\begin{eqnarray}\label{ih}
\frac{\partial x_1}{\partial t} &=& \omega_1 y_1 + D (x_2-x_1) +
\alpha x_1 \nonumber
\\ \nonumber \frac{\partial x_2}{\partial t} &=& \omega_2 y_2+
D (x_1-x_2)+ \alpha x_2 \\ \frac{\partial y_1}{\partial t} &=&
-\omega_1 x_1 + D (y_2 - y_1)+ \alpha y_1 \\ \nonumber \nonumber
\frac{\partial y_2}{\partial t} &=& -\omega_2 x_2 + D (y_1 - y_2) +
\alpha y_2.
\end{eqnarray}

\begin{figure}
\includegraphics[width=10cm,angle=270]{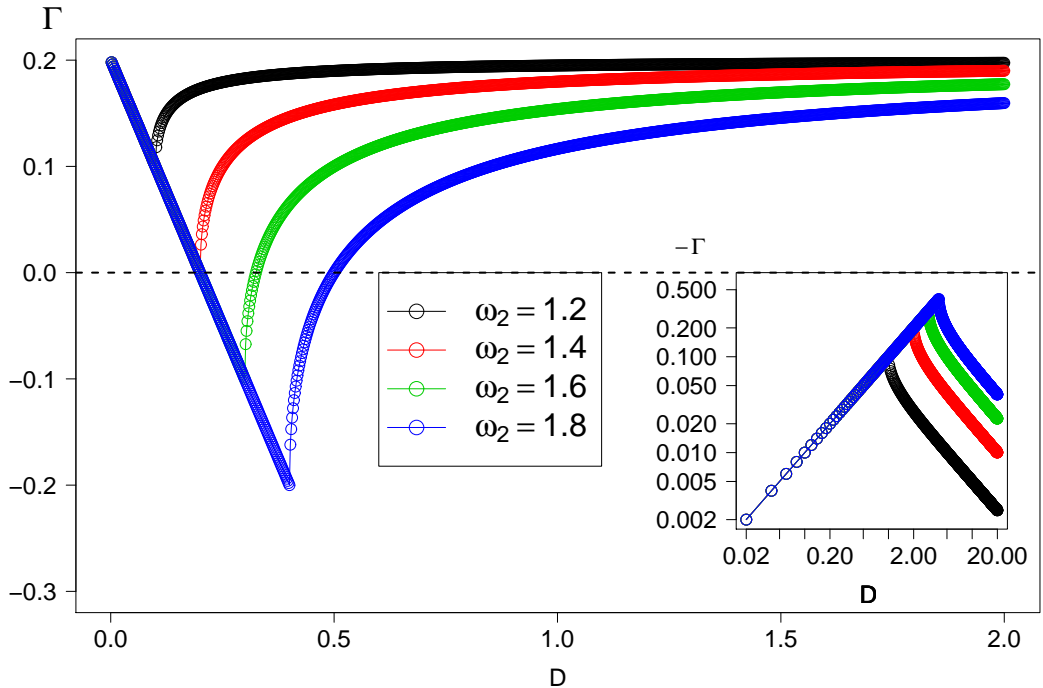}
 \caption{The Lyapunov exponent of the system (\ref{ih}), plotted vs. the migration rate $D$. Parameters are $\omega_1 = 1.$ and $\omega_2 = 1.2,1.4,1.6,1.8$, as
 indicated by the legend; $\alpha = 0.2$. Note that the system attains its maximal stability as $D = \delta/2$, as predicted. The inset shows the same lines for $\alpha = 0$
 in a log-log plot, to emphasize the asymptotic behavior  $D$ and $1/D$. }
 \label{a}
\end{figure}

As this system is linear, it may be diagonalized around the (only)
fixed point at zero. When $\alpha = 0$,  the Lyapunov exponent
$\Gamma$ for that fixed point turns out to be negative as long as
$|\delta| \equiv \omega_2 - \omega_1 \neq 0 $ for any $D$, and
approaches zero (marginal stability) if the dispersal is very small
(no connection between oscillators),  very large (single oscillator
limit) or if the system is homogenous ($\delta \to 0$). The Lyapunov
exponent is given by:
\begin{equation}
\Gamma = \alpha + Re \left[-D+1/2\,\sqrt
{4\,{D}^{2}-4\,\omega_{{1}}\delta+2\,\sqrt {- \left(
\delta+2\,\omega_{{1}} \right) ^{2} \left( -{\delta}^{2}+4\,{D}^{2}
 \right) }-4\,{\omega_{{1}}}^{2}-2\,{\delta}^{2}} \right]
\end{equation}
and its typical behavior is illustrated in Figure \ref{a}. Linearity
implies that if the fixed point is stable it is also globally
attractive. The parametric dependence is characterized by the
following properties:
\begin{itemize}
\item Without loss of generality $\omega_1$ may be scaled to unity
by rescaling of the time. Thus the stability is determined by three
parameters: migration rate, repulsion ($\alpha$) and the
desynchronization term $\delta$.

  \item $\Gamma$ is a nonmonotonic function of the
migration rate; close to zero migration, ($\Gamma - \alpha$)
vanishes linearly with $D$, while for large diffusion it decays like
$1/D$. The optimal dispersal (given that other parameters hold
fixed) is $D = |\delta|/2$; in which case $\Gamma = \alpha -
|\delta|/2$.

  \item  The only effect of $\alpha$ is a rigid upward shift of
  $\Gamma$, as demonstrated in Figure \ref{a}, where the dashed line
  indicates the border between the stable and the unstable regime.

  \item  For fixed migration an increase of $\delta$ always helps
  to stabilize the system, but the effect saturates at  $\delta = 2D$.
  Accordingly, for any $\alpha$, there is a critical diffusion below
  which the system turns to be unstable, independent
  of the level of heterogeneity.
\end{itemize}

The limitations of this toy model are related to its linearity. In
particular, once the  fixed point at zero is unstable, the
oscillation amplitude will grow unboundedly, and no other attractive
manifold is allowed. This should not be the case if the coupled
oscillators are nonlinear. One possible mechanism that may be
responsible for an attractive manifold becomes clear if one looks at
the system (\ref{ih}) in polar coordinates, where $r_i = x_i^2 +
y_i^2$, $\theta_i = arctan(y_i/x_i)$ ($i=1,2$). In this
representation the overall phase $\theta_1+\theta_2$ decouples and
the phase space turns out to be three dimensional,
\begin{eqnarray}\label{3d3}
 \dot{R} &=& \left[ \alpha -2D sin^2 \left( \frac{\phi}{2} \right) \right] R  \nonumber \\
 \dot{r} &=& \left[ \alpha -2Dcos^2\left( \frac{\phi}{2} \right) \right] r   \nonumber  \\
 \dot{\phi} = -2&D&
 \left( \frac{R^2+r^2}{R^2-r^2} \right) sin \phi +
\delta.
\end{eqnarray}
Where the phase desynchronization is $\phi \equiv \theta_2 -
\theta_1$, the amplitude desynchronization is $r \equiv r_2 - r_1$,
and the homogenous manifold is one dimensional $R \equiv r_1+r_2$.
As implied from  (\ref{ih}), the dynamic is  either a flow towards
an attractive  fixed point at $R=0$, or an unbounded growth of $R$,
depending on the relation between the stabilizing desynchronization
factor $\delta$ and the repulsion $\alpha$. If the system is
nonlinear, however, the angular velocity $\omega$ is generally $r$
dependent \cite{Abta_ea2007}. This implies that, although  $\delta$
is too small close to zero,  the "effective $\delta$" may be larger
far from the fixed point, leading to desynchronization and
stabilization.  In such a case one may expect a limit cycle in the
nonlinear system. This, in fact, actually occurs, as will be
demonstrated in the next sections.

Another limitation of the linear model is the fact that the location
of the  fixed point (at zero) is independent of the migration rate.
Equations describing population dynamics support a nontrivial
coexistence fixed point as well,  and diffusion may alter not only
its stability but also its location. In the next section we will
show that the results for the  asymptotic behavior of the Lyapunov
exponent, as well as the saturation of $\Gamma$ for large $\delta$,
may not hold in  more complicated nonlinear dynamics.

\section{Marginally stable Lotka-Volterra dynamics}

Let us now take one step towards more  realistic victim-exploiter
systems, and consider  the Lotka-Volterra continuous  time dynamics,
denoting the density of the predator population by $a$, and the prey
density by $b$.  On a single patch (say, when the population is well
mixed, and the probability of encounter and predation is equal for
any two individuals in the population) the pure LV system is
marginally stable; there is a coexistence fixed point, but the
system may also oscillate with any amplitude around this point, and
the dynamic is determined by the initial conditions. Any type of
stochasticity (demographic, environmental, random migration) will
lead to random wandering of the system between all possible orbits.
The amplitude of oscillations thus performs  some sort of random
walk, with the  average amplitude   growing in time until extinction
\cite{Abta_ea2007}.

To consider the stabilizing effect of spatial differences in the
environment,  we will focus our attention on  two typical
situations: the case of two coupled patches,  where the basic
stability properties are to be  demonstrated, and the case of a one
dimensional "chain" (with $N$ spatial patches and periodic boundary
conditions) where the spatial population profile is examined.

We assume that predator-prey dynamics have the same parameters on
all  sites, the only exception  is the "zero" site where $\sigma$,
the growth rate of the prey, is larger, i.e.,:
\begin{equation} \label{con}
\sigma_n  =  \left\{  \begin{array}{cc}
               \sigma_0 & n=0 \\
               \sigma_1 & else.
             \end{array} \right.
\end{equation}
The corresponding  Lotka-Volterra equations  for a one dimensional
array are:
\begin{eqnarray} \label{LV}
\frac{da_n}{dt} = - \mu a_n + \lambda_A a_n b_n  + D_A(-2 a_n +
a_{n+1} + a_{n-1}) \nonumber \\
\frac{db_n}{dt} = \sigma_n b_n - \lambda_B a_n b_n  + D_B(-2 b_n +
b_{n+1} + b_{n-1}),
\end{eqnarray}
where $\mu$ is the predator death rate, $\lambda$ is the predation
rate and $D_A$ and $D_B$ are the hopping rates of animals from one
spatial patch to the other. Note that one can use the rescaling of
the densities in order to take $\lambda_A = \lambda_B = 1$  and set
$\mu=1$ using the rescaling of time; this is the parametrization
used hereon. Moreover, we focus here on the case of equal
diffusivities $D_A = D_B \equiv D$.

As our system consists of one special site connected to a
"reservoir", it turns out that the character of the  reservoir is
also of importance. A distinction should be made between the case of
a single "oasis" coupled to a desert, (i.e., where $\sigma_1 < 0$ so
the  linear growth rate of both predator and prey is negative on all
the patches except one) and the case where $\sigma_0 > \sigma_1 \geq
0$, where all sites are "active". Hereon, the first case will be
regarded as an "oasis-desert" situation (OD), where the case
$\sigma_1> 0 $ is named the poor-rich (PR) scenario.

\begin{figure}
\includegraphics[width=10cm,angle=270]{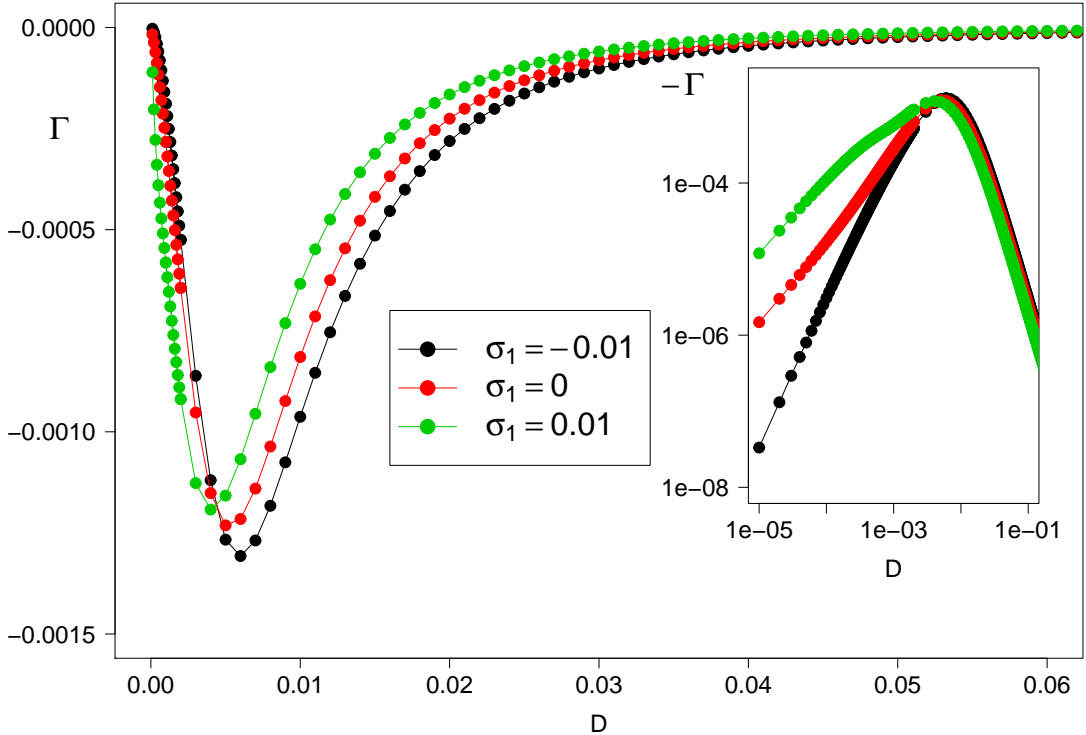}
 \caption{$\Gamma$ vs. $D$ for the  two patch system. The Lyapunov exponent is calculated by numerical diagonalization
 of the stability matrix obtained by linearizing Eqs. (\ref{LV}) around the fixed point,  where the prey fecundity  is  $\sigma_0 = 0.05$ and
  $\sigma_1 = 0.02$, $0$, and $-0.01$, as indicated by the legends. In all cases, the coexistence fixed point is stable for any finite migration.
  The log-log plot (inset) emphasizes the asymptotic behavior in the limits of small and large dispersal: if $\sigma_1 \geq 0$, the linear model predictions
  still hold and $\Gamma \sim D$ at small $D$, while for $\sigma_1 <0$ (coupling of an oasis to a desert) $\Gamma \sim D^2$.
  For large migrations, on the other hand, in all
  cases the asymptote differs from the linear model predictions, and $\Gamma \sim 1/D^3$.}
 \label{fig2}
\end{figure}

\begin{figure}
\includegraphics[width=10cm,angle=270]{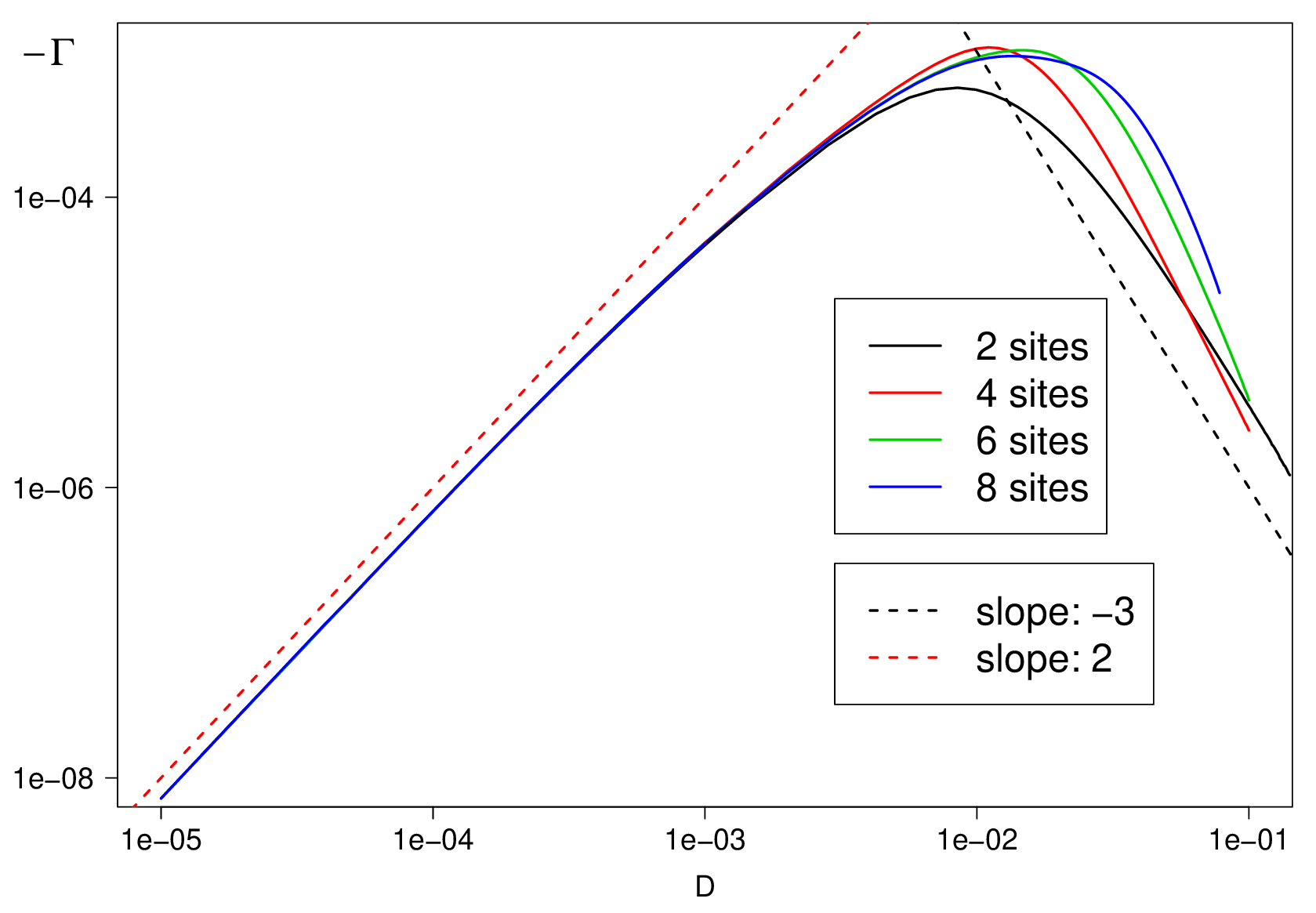}
 \caption{$\Gamma$ vs. $ln(D)$ for two, four, six and eight patch system.
 The Lyapunov exponent is  the result of a  numerical diagonalization
 of the stability matrix obtained by linearizing Eqs. (\ref{LV}) around the fixed point.}
 \label{fig2a}
\end{figure}

To begin, let us consider the two-patch case and examine the
stability properties of the system. As the LV is marginally stable,
one may expect that any spatial heterogeneity will yield stability.
This is, indeed, the situation, as demonstrated in Figures
\ref{fig2}. The general structure is very close to the linear model
predictions with a few exceptions. First, the $D$ dependence of the
Lyapunov exponent at small dispersal holds only in the PR case,
while the OD scenario is characterized by $D^2$ asymptotic. Second,
for the large $D$ asymptotic, $\Gamma$ approaches zero like $1/D^3$
instead of $1/D$ in the linear model. Both asymptotes may be
obtained analytically as explained in appendix A. Note also the
rightward migration of the optimal diffusion as the difference
$\sigma_2 - \sigma_1$ increases; this is the analog of the
$\delta/2$ dependence of the linear model. As previously  explained,
the effects of nonlinearity have to do with the fact that the
location of the fixed point itself depends on the migration rate.
Beyond the two-patch limit, the stability properties are
qualitatively the same, as demonstrated in Figure \ref{fig2a}.

\begin{figure}
 \includegraphics[width=10cm,angle=270]{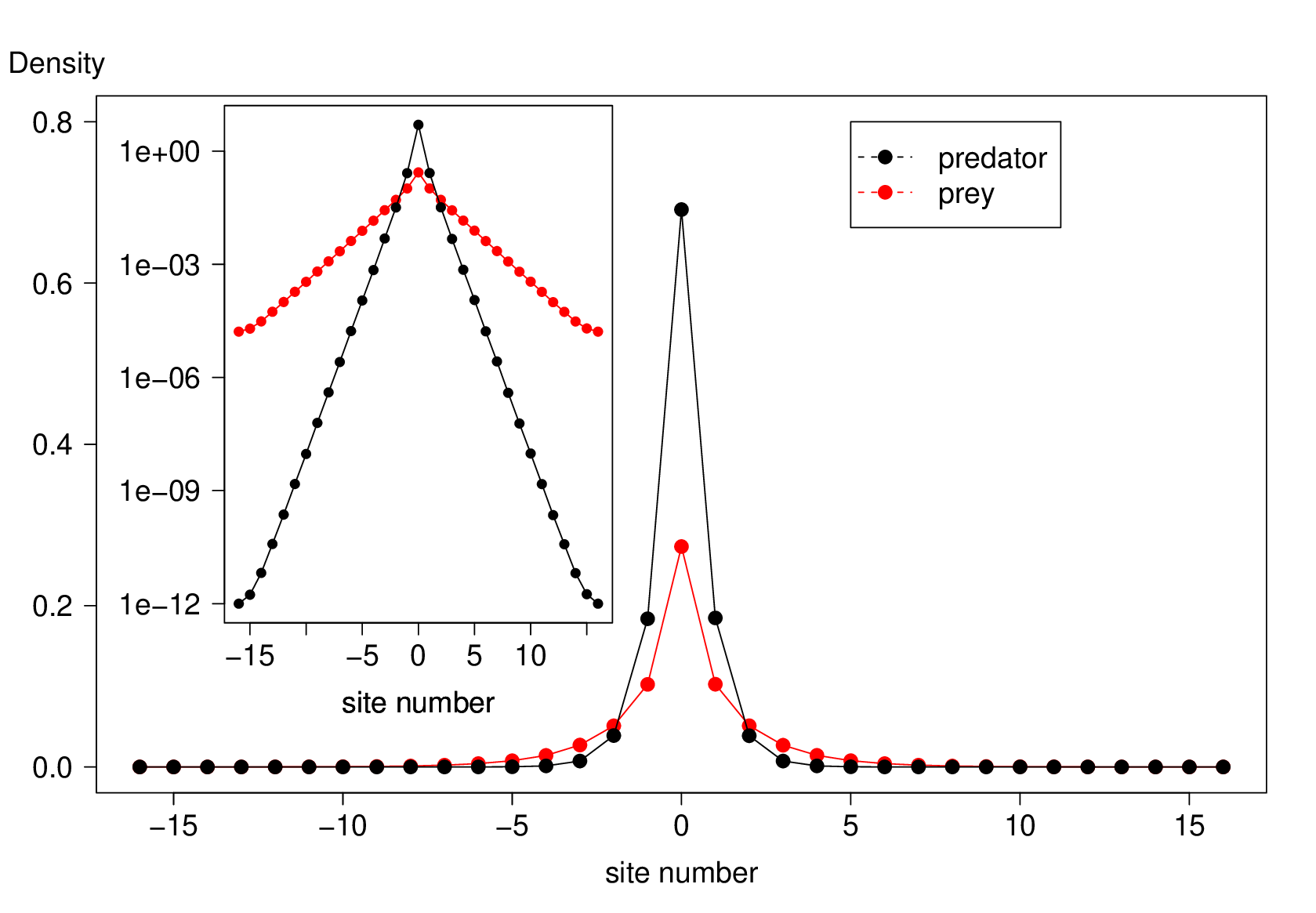}
 \caption{Colony profile for a diffusively coupled oasis-desert  system ($\sigma_0 = 2.5$, $\sigma_1 = -0.25$, $D = 1$).  Note that the ratio between densities
 at the oasis differs from that ratio at the desert.}
 \label{ODspat}
\end{figure}

For the extended system one may consider not only the stability
properties but also the   spatial profile of the population
densities. Let us consider first the oasis-desert scenario.
Numerically solving the spatial configuration for (\ref{LV}) with
the spatial heterogeneity defined by (\ref{con}), one gets the
colony profile presented in  Figure \ref{ODspat}. Clearly, deep in
the desert the populations of both prey and predator are small, so
the nonlinear (predation) terms in (\ref{LV}) are negligible.
Accordingly, each of the species follows asymptotically the profile
of a logistically growing population around an oasis
\cite{Nelson_Shnerb1998, DNS, Kariva_Tilman1997}, and the density
decays like $\exp(-\sigma_1 |x|/D)$ for the prey, and like
$\exp(-\mu |x|/D)$ for the predator. If we adopt the definition of
the size of a colony as the length scale for which the population
density is equal to some constant (this reflects the threshold
introduced by the discreteness of the individuals), this size (up to
logarithmic corrections) is $l_0 D_B / \sigma_1$ for the prey and
$l_0 D_A/ \mu$ for the predator, where $l_0$ is the lattice spacing.

This result has two implications. First, in  case of  multiple
oases, the interesting quantity is the typical distance between them
in units of the colony size; if this is a large number, one should
consider a system of two different patches, while a small number
indicates strong mixing. The chance for a "rescue effect" (where,
due to noise, the population on one patch goes extinct until the
"rescue" by a rare immigrant from another habitat) may be very
different for the victim and for the exploiter, depending on their
death-diffusion ratio in the desert. Second, as demonstrated in
Figure \ref{ODspat}, the spatial decay of the colony may give false
hints regarding the  density at the oasis: in the example presented
here, the predator population is rarely far away, but the predator
abundance is much larger than that of the prey on the oasis. While
the linear death rate of the prey is constant along the system and
determines the exponential decay, its population on the oasis is
dictated by the prey growth rate $\sigma_0$ and is much higher.

\begin{figure}
 \includegraphics[width=10cm,angle=270]{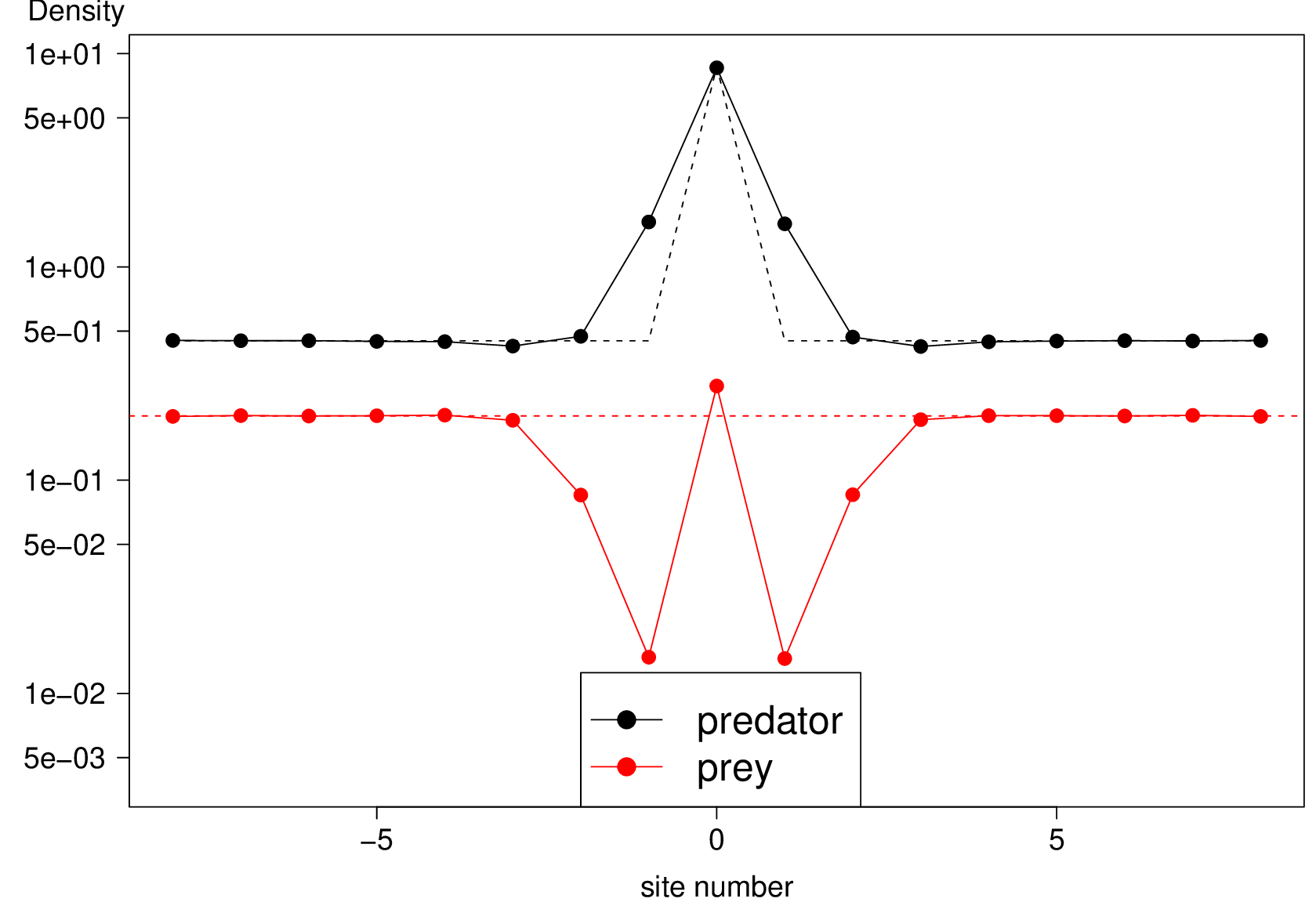}
 \caption{Colony profile for a diffusively coupled poor-rich  system ($\sigma_0 = 43.5$, $\sigma_1 = 27$, ,$D= 0.5$). Without the diffusive coupling, each site supports
 a (marginally stable) fixed point for the predator (dashed black) and for the prey (dashed red). In the diffusively induced stable fixed point, the predator migration
 increases the density of predators in the poor sites close to the rich one, thus depleting the local prey population, and the overall prey profile admits a minimum
 in the neighborhood of the rich site.}
 \label{het}
\end{figure}

The effect of local enrichment becomes even more interesting in the
poor-rich case, as one sees in Figure \ref{het}. An enrichment of
the prey growth rate (say, by increasing the food supply) on a site
leads to  \emph{depletion} of  the prey in its neighborhood, while
leaving the population on the rich site above its pre enrichment
level. Surprisingly, the behavior of the prey density is
paradoxical, but the "paradox" is nonlocal.

\section{Unstable systems: Nicholson-Bailey like model}

The original mathematical description of a host-parasitoid  system
was formulated by Nicholson and Bailey \cite{Nicholson_Bailey1935}.
If the host density is given by $H$ and the parasitoid is $P$,
Nicholson and Bailey map for nonoverlapping generations is:
\begin{eqnarray} \label{NB0}
H_{t+1} &=& q H_t e^{-z P_t} \nonumber \\
P_{t+1} &=& c H_t (1-e^{-z P_t}),
\end{eqnarray}
where $q$ is the fecundity of the host in the absence of the
parasitoid, $z$ is the infectivity (the chance of a single
parasitoid to infect the host) and $c$ is the number of parasites
produced by a single infection.

The essential feature of the NB model is its instability - it allows
ever growing oscillations around its coexistence fixed point.
However, the model does not allow explicitly for extinction, as P
and H are positive definite along the process. As the oscillations
grow, the minimal distance from the extinction phase for each of the
species decays rapidly. Hence, for any realistic system, where small
noise or the discreteness of individuals are taken into account, the
system flows deterministically to the extinction phase.

\begin{figure}
 \includegraphics[width=10cm,angle=270]{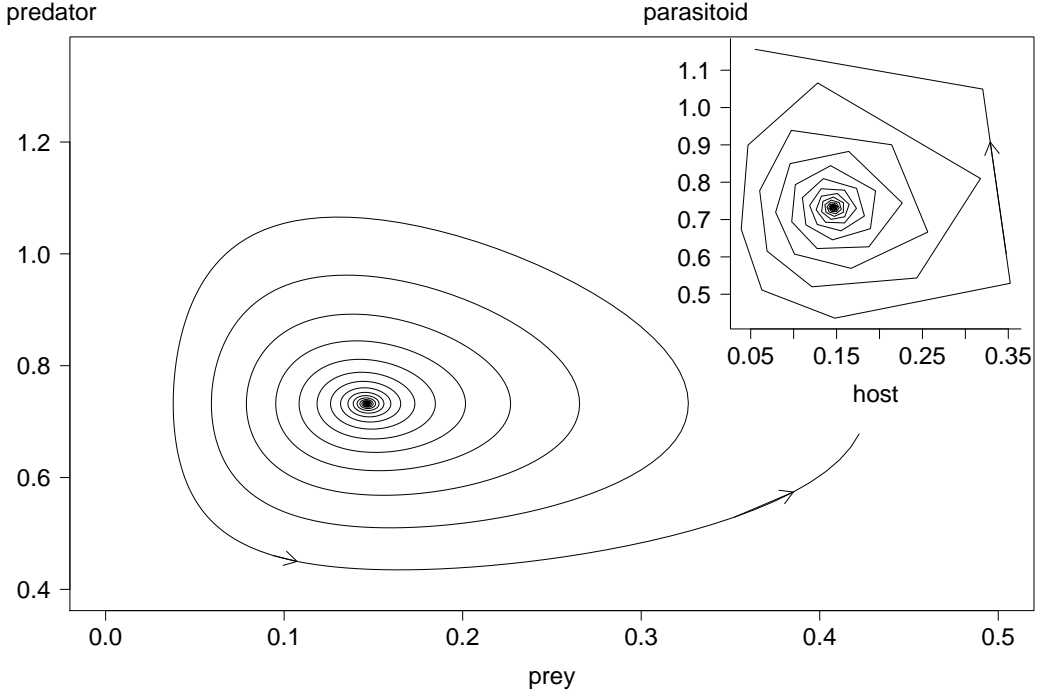}
 \caption{Phase portrait, calculated by  numerical integration of  Eqs. \ref{NB}, shows the main characteristics of the
 NB continuous analogue: the instability of the fixed point, the exponential growth of the oscillations, and the fact that the
 trajectory cannot cross the boundaries of zero population for either predator or prey. The corresponding trajectories of the map
 (\ref{NB0}) for host-parasitoid are shown in the inset.}
 \label{exp}
\end{figure}

One technical characteristic of the NB dynamics makes its analysis
problematic in the context of the current discussion. Eqs.
(\ref{NB0}) assume nonoverlapping generations, and the corresponding
map, when extended to diffusively coupled spatial domains, supports
attractive periodic orbits with finite basin of attraction
\cite{adler}. Even in the absence of spatial inhomogeneities some
initial conditions are trapped in the periodic orbits, and it is
hard to distinguish between this effect and stabilization due to
spatial differences. In order to focus the discussion on the
consequences of local enrichment, we will switch here to  continuous
time dynamics that  imitates the relevant features of the NB model.
For the sake of nonlinear dynamics analysis considered here, our
model is equivalent to the  small $z$ limit of the NB system, where
the map may be approximated by continuous time equations, as
emphasized in figure \ref{exp}.

Let us assume, thus,  that in a predator-prey system the encounter
between a single predator and prey may result  in predation, but the
predation probability increases if two predators encounter a single
prey (i.e., there is a predation "Allee effect").  We "enrich" the
LV system with additional terms that correspond to  this ally
predation process:

\begin{eqnarray} \label{NB}
\frac{da}{dt} = - \mu a + \lambda_A a b + \alpha a^2 b  \nonumber \\
\frac{db}{dt} = \sigma b - \lambda_B a b -  \alpha a^2 b.
\end{eqnarray}
The resulting equations admit a coexistence fixed point at $a =
(\Delta - \lambda) / 2 \alpha $, $\ b = 2 \mu / (\Delta + \lambda)$,
where $\Delta = \sqrt{\lambda^2 + 4 \alpha \sigma}$. This fixed
point is unstable, with a positive local Lyapunov exponent. The
amplitude of oscillation increases rapidly with time, but the system
cannot reach the extinction phase unless some noise is presented.

To include spatial structure, we switch to a nondimensionalized form
of the diffusively coupled patch system, where the definition of
constants is identical with Eq. \ref{LV}:

\begin{eqnarray} \label{NB1}
\frac{da_n}{dt} = - a_n +  a_n b_n + \alpha a_n^2 b_n + D_A(-2 a_n +
a_{n+1} + a_{n-1}) \nonumber \\
\frac{db_n}{dt} = \sigma_n b_n - a_n b_n -  \alpha a_n^2 b_n +
D_B(-2 b_n + b_{n+1} + b_{n-1}).
\end{eqnarray}
Again, one should expect to recover the zero dimensional dynamics as
$D$ approaches zero and infinity. In between, the resulting dynamics
turn out to be quite rich. As in the LV case, we will check the
two-patch system  to get the general picture for the stability of
the system, then look at a multi-patch example to see the spatial
profile.

\begin{figure}
 \includegraphics[width=10cm,angle=270]{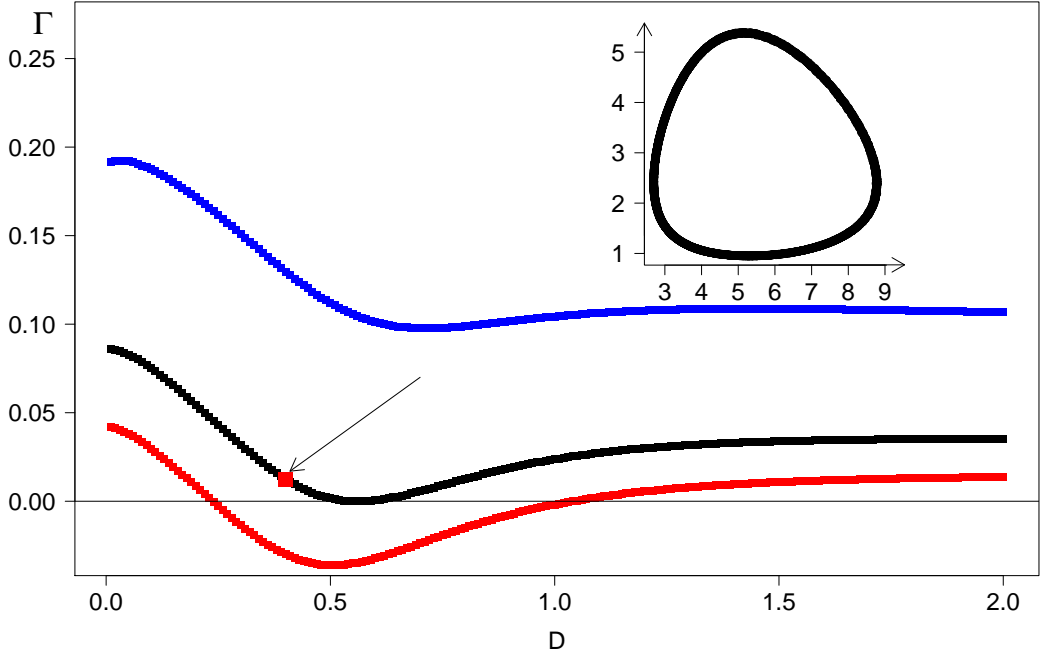}
 \caption{Stability diagram for the coexistence fixed point of the two-patch Nicholson-Bailey like
 system, Eqs. (\ref{NB1}). Parameters are $\sigma_0 = 1$, $\sigma_1 = -0.1$, and $\alpha = 0.1$ (low) $\alpha = 0.4$ (middle) and
 $\alpha = 1$ (upper line). Only the low $\alpha$ allows for a finite region of dispersal rate in which a stable coexistence fixed point is supported.
 However, even in the unstable region the system may flow into a limit cycle   or other attractive manifolds, see Appendix B. The inset shows a stable orbit
 (projected on the homogenous manifold) for
 the system parameters at the  point indicated by an arrow. }
 \label{st}
\end{figure}

The linear  model admits two options:  in the presence of spacial
variation the unstable system may remain unstable (in which case it
flows to extinction exponentially fast) or admit (for intermediate
diffusivities) an attractive fixed point. As exemplified in Figure
\ref{st}, the nonlinear model (\ref{NB1}) turns out to admit another
option:  while the fixed point may be unstable,  the system does not
flow to extinction but to an attractive manifold like a limit cycle.
This gives rise to many complications in the system behavior,
including various types of bifurcations, bistability and hysteresis.
Partial analysis of the bifurcations and the phase portraits for the
unstable case is presented in Appendix B.

\begin{figure}
 \includegraphics[width=10cm,angle=270]{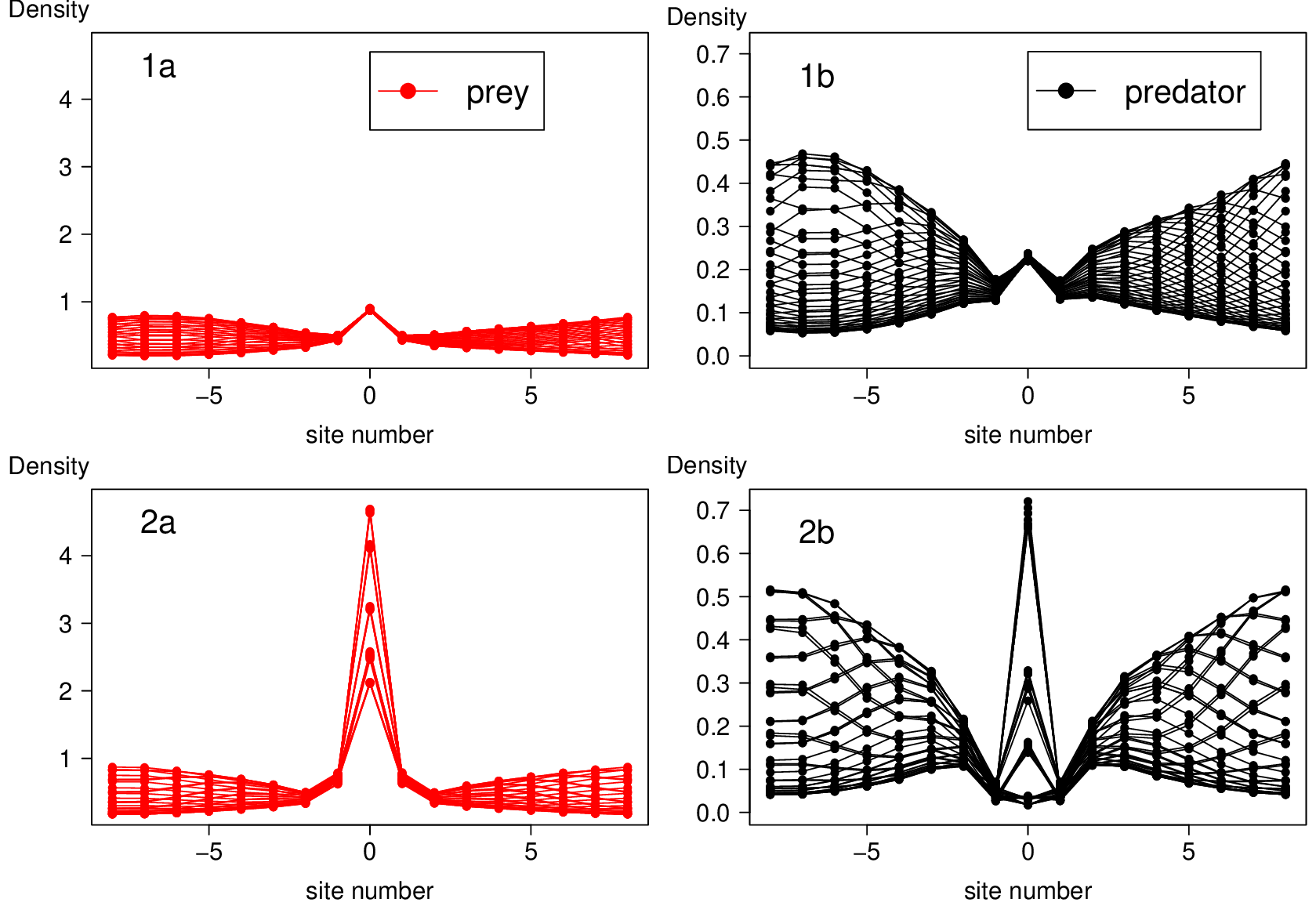}
 \caption{Population profile for a 16 site system with periodic boundary conditions  (a few snapshots taken along the evolution of the system) for the
 prey (red) and the predator (black).
 The two upper panels correspond to $\alpha = 0.02$,  $\sigma_0 = 5$, $\sigma_1 = 2.25$ and $D=0.5$. One notices that the enrichment almost fixes the population on the rich (zero) site at
 the middle, while in the far zone  (the "wings") the system oscillates. The lower panels show what happens as the spatial variation is increased: here
 $\sigma_0 = 21.5$ and all other
 parameters are the same. Due to the increase of heterogeneity, cycles also developed on the rich site, while in the sites close to the origin, their amplitude
 remains
 negligible.   }
 \label{PR_NLE}
\end{figure}

\begin{figure}
 \includegraphics[width=10cm,angle=270]{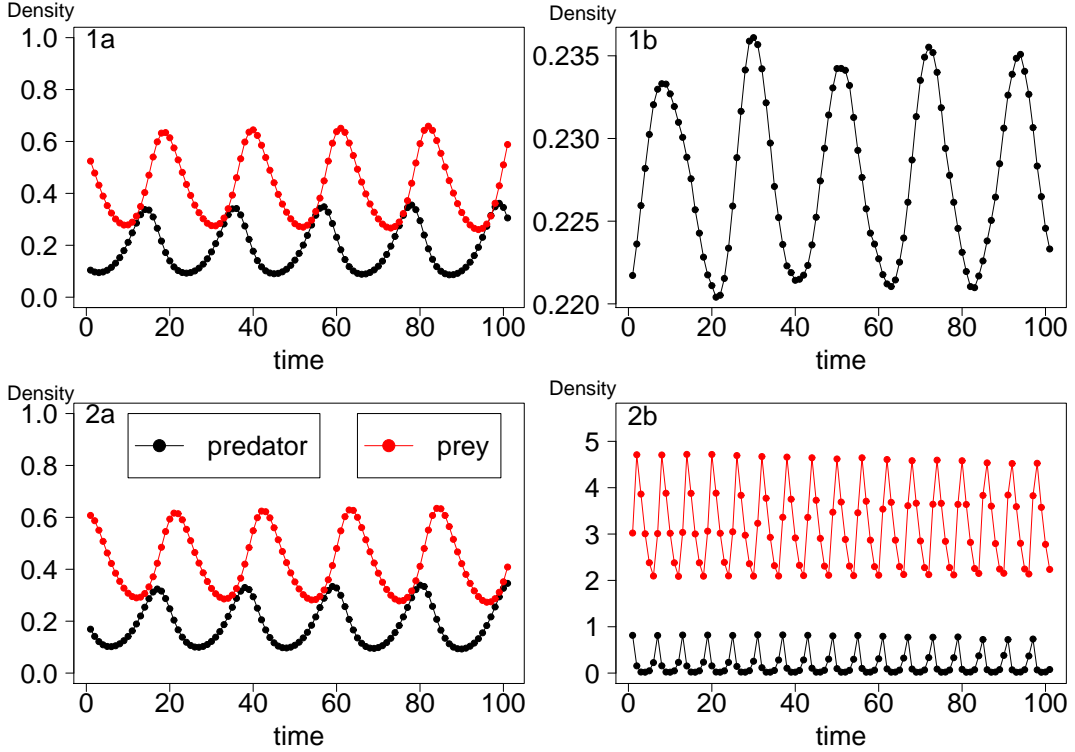}
 \caption{Prey (red) and predator (black) populations vs. time for $\sigma_0 = 5$ at the "wings" (far zone - see figure \ref{PR_NLE}) (panel 1a) and at the
 rich point (panel 1b; the prey at the enriched point is not presented as its oscillations are relatively small). Panel  2a shows the prey and
 predator
 at the wings for $\sigma_0 = 21.5$, while in  2b the oscillations at the rich point are graphed for the same heterogeneity.       }
 \label{ost}
\end{figure}

\begin{figure}
 \includegraphics[width=10cm,angle=270]{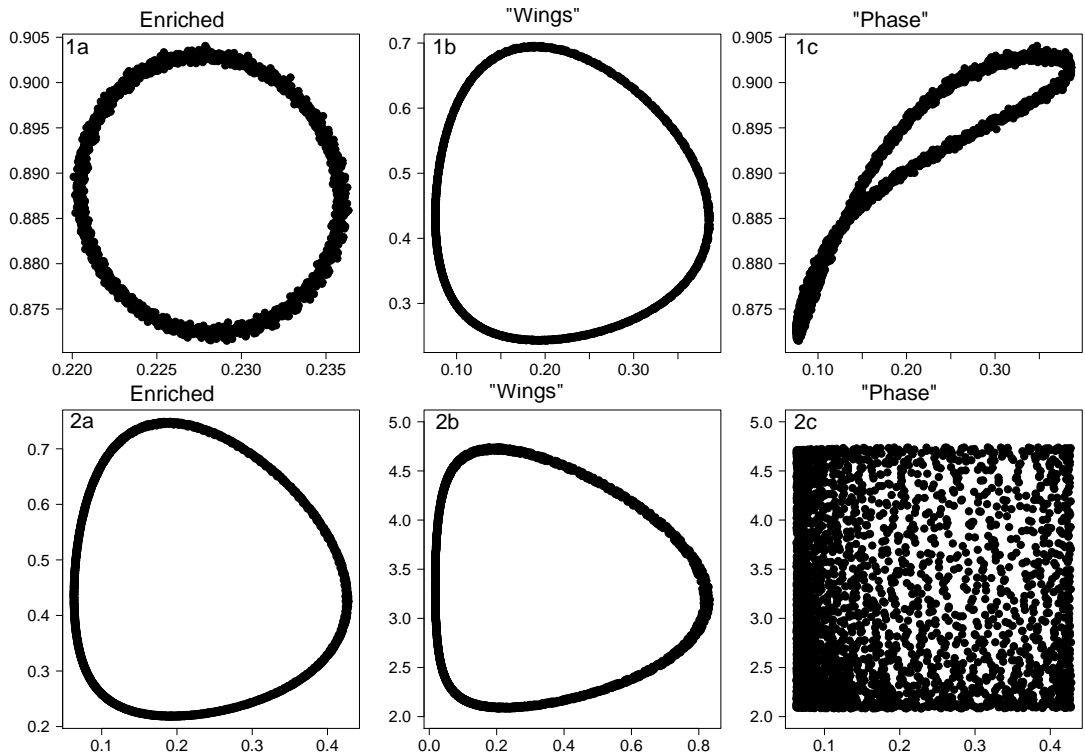}
 \caption{Phase portrait of the populations (predator vs. prey) for the cases considered in figures \ref{PR_NLE} and \ref{ost}.
 The $\sigma_0 = 5$ case
 yields the upper panels, where 1a is the phase portrait at a point on the wing, 1b is on the enriched site (not the low amplitude of oscillations) and
 in 1c  the prey at the center is graphed  vs. the predator on the wings. 2a and 2b are the same graphs  for the $\sigma_0=21.5$ case,
 and the fact that the oscillations are incommensurate
 manifests itself in panel 2c. In all subfigures the $y$ axis is the predator density and the x axis is the prey density.}
 \label{torus}
\end{figure}

\begin{figure}
 \includegraphics[width=10cm,angle=270]{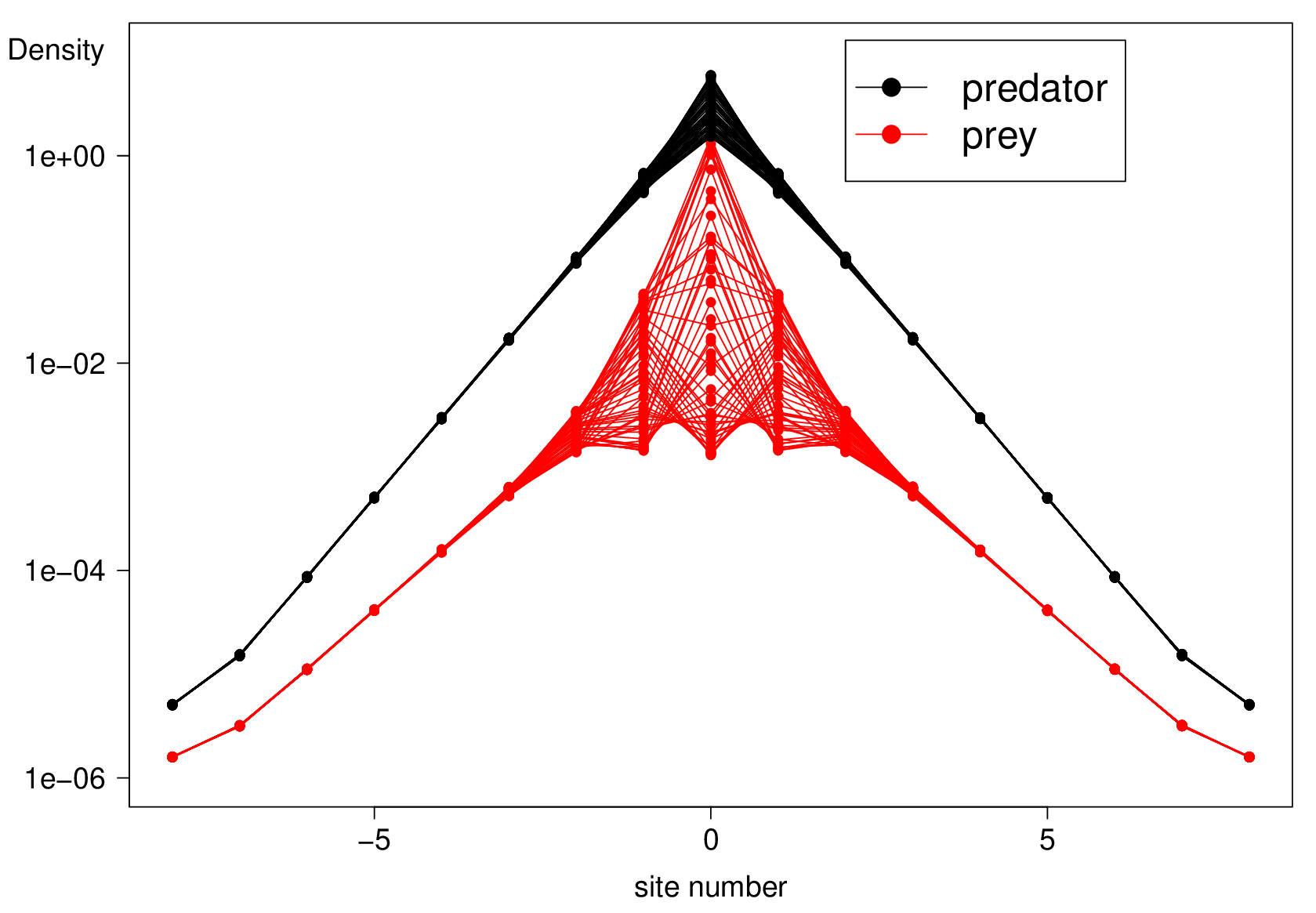}
 \caption{Prey (red) and predator(black) populations vs. time for $\sigma_0 = 21.5$, $\sigma_1 = -0.5$, $\alpha = 0.02$ and $D = 0.5$. Here the oasis (at zero) is
 coupled to a desert, and hence the oscillations take place only at  the enriched location.       }
 \label{OD_NLE}
\end{figure}

As in the marginally stable LV case, one observes a nonlocal effect
of enrichment on the spatial profile. In  Figure \ref{PR_NLE}  a few
possibilities are demonstrated. It may happen that nutrient
enrichment  at a single site causes the other parts of the system to
change periodically, while the rich site population is almost fixed.
Increasing the spatial variation, the amplitude of oscillations on
the rich site  grow (Figure \ref{ost}), and at the same time the
phase locking with the "wings" is lost.  At the end, the
oscillations at the center and on the wings become incommensurate,
giving rise to an attractive torus, as demonstrated in  Figure
\ref{torus}. In the oasis-desert case the effect is less dramatic,
and the oscillations appear first in the enriched region (Figure
\ref{OD_NLE}).

\section{Discussion}

The effect of quenched spatial variations, where the habitat is made
of local patches connected by dispersal, is stabilizing.  Besides
the survey of the possible states of this dynamical system and the
asymptotical behavior for large and small migration rates, the
nonlocal effect of enrichment  emerges from the numerics as a new
feature. In particular, it turns out that local enrichment may
stabilize otherwise unstable system, and that the limit cycle may
involve periods of large amplitude far from the enriched site, while
the rich site itself involves cycles of negligible amplitude.
Increasing the spatial heterogeneity, one observes an increase in
oscillation amplitude on the rich site, while its neighborhood
remains relatively calm.  The oscillations at the enriched  site may
be incommensurate with the period of oscillations at the "wings",
such that  the system flows into a stable torus instead of  a limit
cycle.

This observation may solve the "paradox of enrichment"
\cite{Rosenzweig1971} in some cases. Many victim-exploiter models
predict an increase in oscillation amplitude as a result of
increased prey growth rate or carrying capacity. The empirical
support of this prediction, however, is quite limited
\cite{Tilman_Wedin1991} and in many experiments the effect is absent
\cite{Murdoch_McCauley1985,Kirk1998}. From the results here one
concludes that in some cases, small spatial fluctuations in the
enrichment, e.g., small inhomogeneities in the concentration of
food, sunlight or other resources, may stabilize the system and
avoid the "paradoxical" behavior.

What degree of enrichment uniformity  is required  in order to
observe the paradoxical behavior? Such a question may be analyzed
very easily using the toy model presented here as  a basic tool for
parameter estimation. As stressed above, only three parameters
control the system close to the fixed point: the dispersal rate $D$,
the repulsion $\alpha$ and the desynchronization factor $\delta$.
The threshold between  a "uniform like" and heterogenous system
appears at $D \sim \delta$, i.e., when the migration rate in/out of
a patch (here a patch is defined as a region where the resource
distribution is uniform) is close to the desynchronazation rate,
given by the difference in oscillation period  between patches. If
diffusion is much larger, the system should be considered as
uniform. Much smaller diffusion corresponds to the situation of
almost independent patches.

For an experimental system, one should try to gain a rough estimate
of the above parameters in order to ascertain the level of
enrichment needed to yield a strong enough effect. Here we exemplify
these considerations using  two experiments: one on a predator-prey
system and the other  host-parasitoid.

In the experiment of Kerr et al \cite{Kerr_ea2006}, E-coli bacteria
is the host and its viral pathogen, T4 coliphage,  is the parasite.
In the well-mixed case, the system survives about 24h before the
bacteria undergoes extinction; this corresponds to  $\alpha  \sim
10^{-5}$ seconds. The migration in that experiment is controlled by
the experimentalists, as a robot taking biological material between
otherwise disconnected patches. The authors present simulations,
based on empirically calibrated cellular automata, that predict
oscillations with a time scale of about 10 days, so $\omega_1$ is
about $10^{-5}$, i.e.,  close to $\alpha$. If one chooses a day as
the unit time, $\omega \sim \alpha \sim 1$. From the linear system
analysis, it seems that this system should be coupled to another one
with $\omega_2 \sim 2 \omega_1$ in order to observe stabilization
due to spatial heterogeneity.

Another example is the experiment of Holyoak and Lawler
\cite{Holyoak_Lawler1996}, where the predaceous ciliate
\emph{Didinium nasutom} is feeding on the bacterivorous  ciliate
\emph{Colpidium cf. Striatum}. The diffusion constant for these
small (length of order 0.1-0.05 mm)   creatures  depends on the
level of water turbulence and on the size of the tubes connecting
different microcosms. The extinction time in the well mixed case is
about 70 days ($\alpha \sim 5*10^{-6}$ seconds) and $\omega_1$ is
about 10 days. In the appropriate dispersal, one may observe
stabilization if the system is coupled to another set of microcosms
with $\delta = \omega_2 - \omega_1 \sim 0.05$, i.e., relatively
small enrichment will allow for stabilization.

We acknowledge helpful discussions with Marcel Holyoak. This work
was supported by the  EU 6th framework CO3 pathfinder and DAPHNet.

\section{Appendix A}

In this appendix we discuss the dependence of the system's stability
(its Lyapunov exponent) on the rate of animal dispersal. As
explained above, one should recover the mean field results in the
case of no migration, $D=0$,  and as the migration rate approaches
infinity. For the Lotka-Voltera, marginally stable case, the
Lyapunov exponent  vanishes in these two limits, while in between it
is finite and negative. We now want to explore the two limits more
carefully and attain the asymptotic dependence of the Lyapunov
exponent in terms of the system parameters.

To begin, let us consider the simplest spatially explicit case,
i.e., a two-patch LV system. The dynamics in such a  case are four
dimensional and  described by:
\begin{eqnarray} \label{LV_2p}
\frac{da_1}{dt} = -  a_1 +  a_1 b_1  + D( a_2 - a_1) \nonumber \\
\frac{da_2}{dt} = -  a_2 +  a_2 b_2  + D( a_1 - a_2) \nonumber \\
\frac{db_1}{dt} = \sigma_1 b_1 -  a_1 b_1  + D( b_2 - b_1) \nonumber \\
\frac{db_2}{dt} = \sigma_2 b_2 -  a_2 b_2  + D( b_1 - b_2),
\end{eqnarray}
where the growth parameter  $\sigma$ is heterogenous. Clearly, as
$D$ approaches infinity the differences between patches, $\rho
\equiv a_1 - a_2$ and $\theta \equiv b_1 - b_2$ approach zero, so it
is useful to   rotate the coordinate system in order to separate
between the homogenous manifold ($A = (a_1 + a_2)/2$ and $B = (b_1 +
b_2)/2$) and the $\rho - \theta$ manifold. Defining, now, $\epsilon
= 1/D$, and assuming that $A = A_0 + \epsilon A_1 ...$ (and the same
for B) while $\rho = \epsilon \rho_1 + \epsilon^2 \rho_2 ...$ (and
the same for $\theta$), one can solve for the fixed point of the set
of equations (\ref{LV_2p}),  order by order in $\epsilon$ up to,
say, order $n$. This solution is then plugged into the stability
matrix, and the roots of the characteristic polynomial may then be
found (again, order by order in $\epsilon$) up to ${\cal O}(n)$.
With that, one finds that the leading contribution to the Lyapunov
exponent (the first order for which the root of the characteristic
polynomial admits a real part) is $n=3$. The asymptotics of the
Lyapunov exponent for large D turns out to be:
\begin{equation}
\Gamma = - \frac{(\sigma_1 - \sigma_2)^2 (\sigma_1+\sigma_2)}{64
D^3}.
\end{equation}

Close to $D=0$, on the other hand, one should use another technique.
Solving $a_1,a_2,b_1$ and $b_2$ to the n-th order in $D$, these have
to be plugged into the stability matrix, from which the
characteristic polynomial is extracted. For $D=0$, two (in the
oasis-desert case) or all four (in the poor-rich case)
 of the eigenvalues are purely imaginary. The
 corrections to these imaginary eigenvalues are written as a power
 series in $D$, and one finds that the real part of the slowly
 decaying eigenfunction (i.e., the Lyapunov exponent) is
 ${\cal O}(D)$ in the poor-rich case, but only  ${\cal O}(D^2)$
 for the oasis-desert situation. For $\sigma_2 < 0$ (oasis
 desert situation) one gets:
\begin{equation}
\Gamma = -{\frac {{{ \sigma_2}}^{3}+{ \sigma_2}\,{ \sigma_1}+{{
\sigma_1}}^{2}+{ \sigma_1}}{{ \sigma_2}\, \left( { \sigma_1}\,{{
\sigma_2}}^{2}+{{ \sigma_2}}^{2}+{{ \sigma_1}}^{2}+{ \sigma_1}
 \right) }} \frac{D^2}{2}.
\end{equation}
While in the poor-rich case for $\sigma_1 > \sigma_2$ the exponent
is:
\begin{equation}
\Gamma = - \frac{\sigma_1 + \sigma_2}{2 \sigma_1} D.
\end{equation}
This result does not hold when the system approaches the uniform
limit $\sigma_1 = \sigma_2$, as it has been obtained perturbatively,
neglecting the degeneracy appearing for these parameters.

\section{ Appendix B}

In this appendix we consider the various equilibrium states of the
unstable dynamics  (\ref{NB}) on a heterogenous environment
(\ref{NB1}). The 4d system is quite complicated and allows for many
types of bifurcations. In the following, a general sketch of the
main types of dynamical behavior is given.

\begin{figure}
 \includegraphics[width=10cm,angle=270]{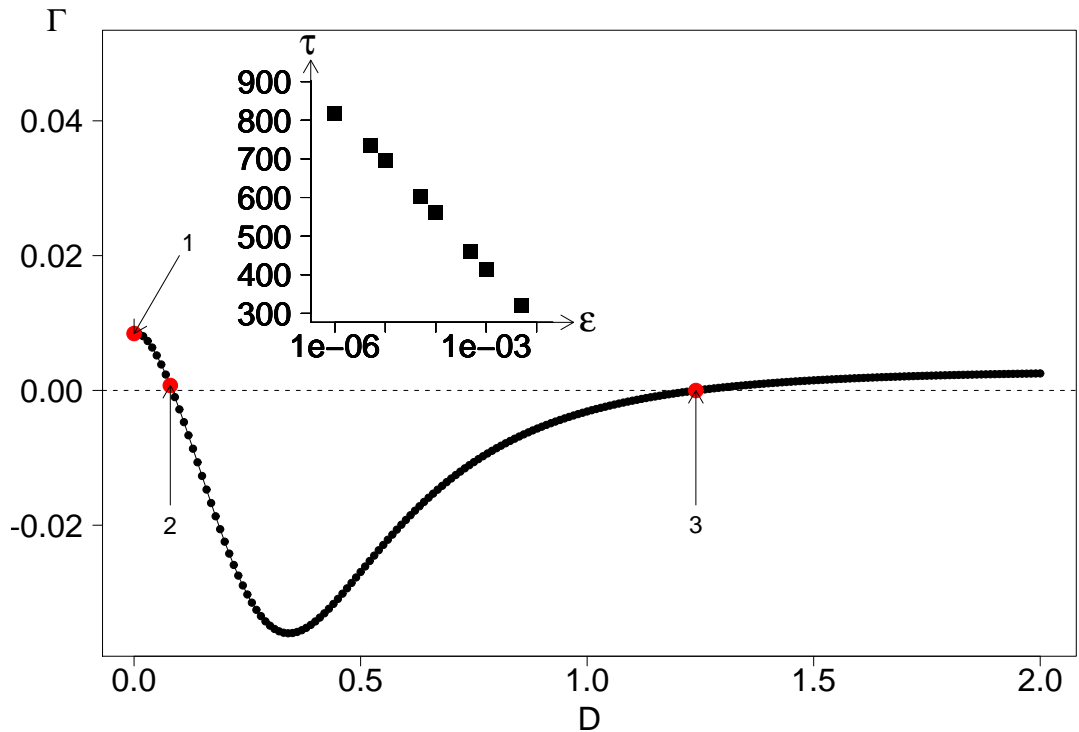}
 \caption{Nicholson Bailey system (\ref{NB1}), where the parameters (specified in Figure \ref{st}) allow for a locally stable coexistence point.
 The bifurcation takes place at the points indicated by 2 and 3, and between 2 and 1 a limit cycle exists. This limit cycle annihilates with the saddle
 via homoclinic bifurcation, characterized by a logarithmically diverging period of oscillations, as indicated in the inset.  For $D$ values right  above the upper
 point 3 we have observed trajectories that converge to a limit cycle, but this cycle disappears close to that point and leaves an unstable system.}
 \label{NB1f}
\end{figure}

Let us take a look at Figure \ref{NB1f}, which describes the
Lyapunov exponent of the coexistence fixed point vs. $D$. In the
region of intermediate migration  (between 2 and 3), the coexistence
point becomes  attractive due to the heterogeneity, similar to the
LV system. However, at 2  the system undergoes a supercritical Hopf
bifurcation, and a globally attractive limit cycle appears. As the
diffusion constant approaches zero the radius of this limit cycle
diverges, and finally it losses its stability via homoclinic
bifurcation, and the cycle period diverges logarithmically as shown
in the inset to Figure \ref{NB1f}.

\begin{figure}
 \includegraphics[width=10cm,angle=270]{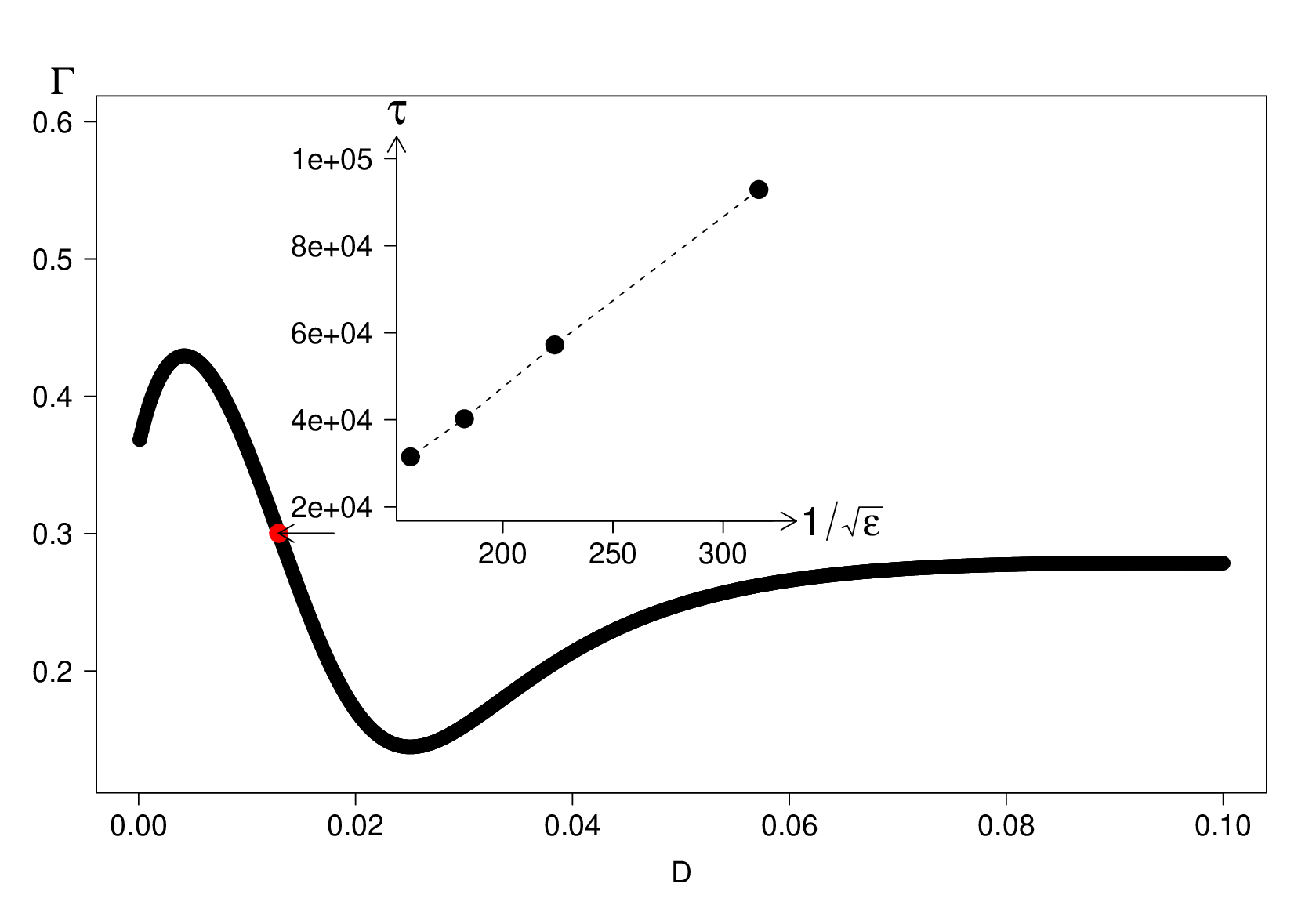}
 \caption{Nicholson-Bailey like system with large $\alpha$. Here there is no stable fixed point for any value of migration, yet
 it may support a limit cycle. This limit cycle exists only to the left of the dot indicated by the arrow. At this point the limit cycle disappears
 in an infinite period bifurcation. The  time spent by the system in its unstable phase (to the right of the indicated point) close to "ghost" of the limit
 cycle diverges close to the bifurcation like $1/\sqrt{\epsilon}$, where $\epsilon \equiv D-D_c$ is the distance from the bifurcation point (inset).   }
 \label{NB2}
\end{figure}

There exists, however, another set of parameters, where the
coexistence fixed point does not acquire stability for any diffusion
rate, as demonstrated in Figure \ref{NB2}. It turns out that, in
this case, the limit cycle born via homoclinic bifurcation at $D=0$
loses its stability through an infinite period bifurcation  at
larger dispersal values, leaving an unstable system with
oscillations that grow unboundedly until extinction. The stable
cycle reemerges at even higher values of $D$, while now its size
grows as diffusion increases. It should be noted that in some cases
(like the parameter range that corresponds to the middle curve in
Figure \ref{st}) we have observed a limit cycle behavior for any
value of $D$.

\newpage
\bibliographystyle{unsrt}
\bibliography{bib}

\end{document}